\documentclass[aps,prl,twocolumn,groupedaddress,showpacs,showkeys]{revtex4-1}
\usepackage{verbatim}
\usepackage{amsmath}
\usepackage{graphicx}
\usepackage{array}
\usepackage{SIunits}
\usepackage{color}
\usepackage{float}
\usepackage{placeins}
\usepackage{epstopdf}
\usepackage{upgreek}
\usepackage{booktabs}
\usepackage{longtable}
\usepackage{calrsfs}
\newcolumntype{P}[1]{>{\centering\arraybackslash}p{#1}}
\newcommand{\pryso}{Pr$^{3+}$:Y$_2$SiO$_5$}

\newcommand{\eryso}{Er$^{3+}$:Y$_2$SiO$_5$}
\newcommand{\erlyf}{Er$^{3+}:^7$LiYF$_4$}

\newcommand{\dhdhhh}{$^4\!I_{13/2}-\,^4I_{15/2}\,$}
\newcommand{\pumping}{$^4\!I_{11/2}-\,^4I_{13/2}\,$}
\newcommand{\Lb}{\mathcal{L}}

\begin{document}
%\preprint{}

\title{A naturally trapped rare-earth doped solid-state superradiant laser clock}


%\author{Mahmood Sabooni$^{a,b}$, YYY$^{a}$, ZZZ$^{c}$}
\author{Mahmood Sabooni}
%\email[]{msabooni@uwaterloo.ca}
%\thanks{}

\affiliation{Institute for Quantum Computing, Department of Physics and Astronomy, University of Waterloo, Waterloo, Ontario, N2L 3G1, Canada.}
%\affiliation{$^{a}$ Department of Physics, Lund University, P.O.~Box 118, SE-22100 Lund, Sweden}
%\affiliation{???}%$^{a}$ Department of Physics, University of Tehran, Kargar Shomali Avenue, Tehran 14399-55961, Iran}
%\affiliation{$^{c}$ Department of Electroscience Electromagnetic Theory, Lund University, P.O.~Box 118, SE-22100 Lund, Sweden}
%\affiliation{**Spectrum Lab, P. O. Box 173510, Montana State University, Bozeman, Montana 59717}


%\date{\today}

\begin{abstract}
\par
We propose a solid-state based superradiance laser which is almost insensitive to the cavity mirror vibration. Therefore, it can compete with the best frequency-stable local oscillators. The long coherence time and the large optical density of rare-earth-ions (REIs) doped solids are employed to find a regime to demonstrate a steady-state laser emission with linewidth smaller than the atomic decay rate. The experimental parameters are discussed and intracavity photon number and laser linewidth are calculated based on the mean-field theory. A procedure for measuring absolute laser linewidth is proposed.
\end{abstract}

\pacs{42.50.Ct, 03.67.Hk, 42.50.Gy, 42.50.Md}
%\keywords{}

\maketitle

%\section{Introduction}

%\section{Introduction\label{sec:Intro}}

\par
%\textbf{Introduction:}
The frequency reference improvement is a vital step towards progress in a wide range of applications in precision metrology, fundamental tests in the quantum information science, and quantum optics up to technology related applications, such as communication and navigation systems. Optical atomic clock precision and stability are limited to the frequency-stable laser local oscillators (LLOs) \cite{Campbell2017, Schioppo2017, Ludlow2015, Bishof2013}. The main obstacle against improving the frequency-stability of LLOs is the thermal noise in the optical cavity length which is already in the order of the size of a single proton ($\sim10^{-15}m$) \cite{Thorpe2011}. This is because of the frequency stability $\frac{\Delta \nu}{\nu}$ directly proportional to $\frac{\Delta L}{L}$. %

\begin{figure}[ht]
 \centering
    \includegraphics[width=8cm]{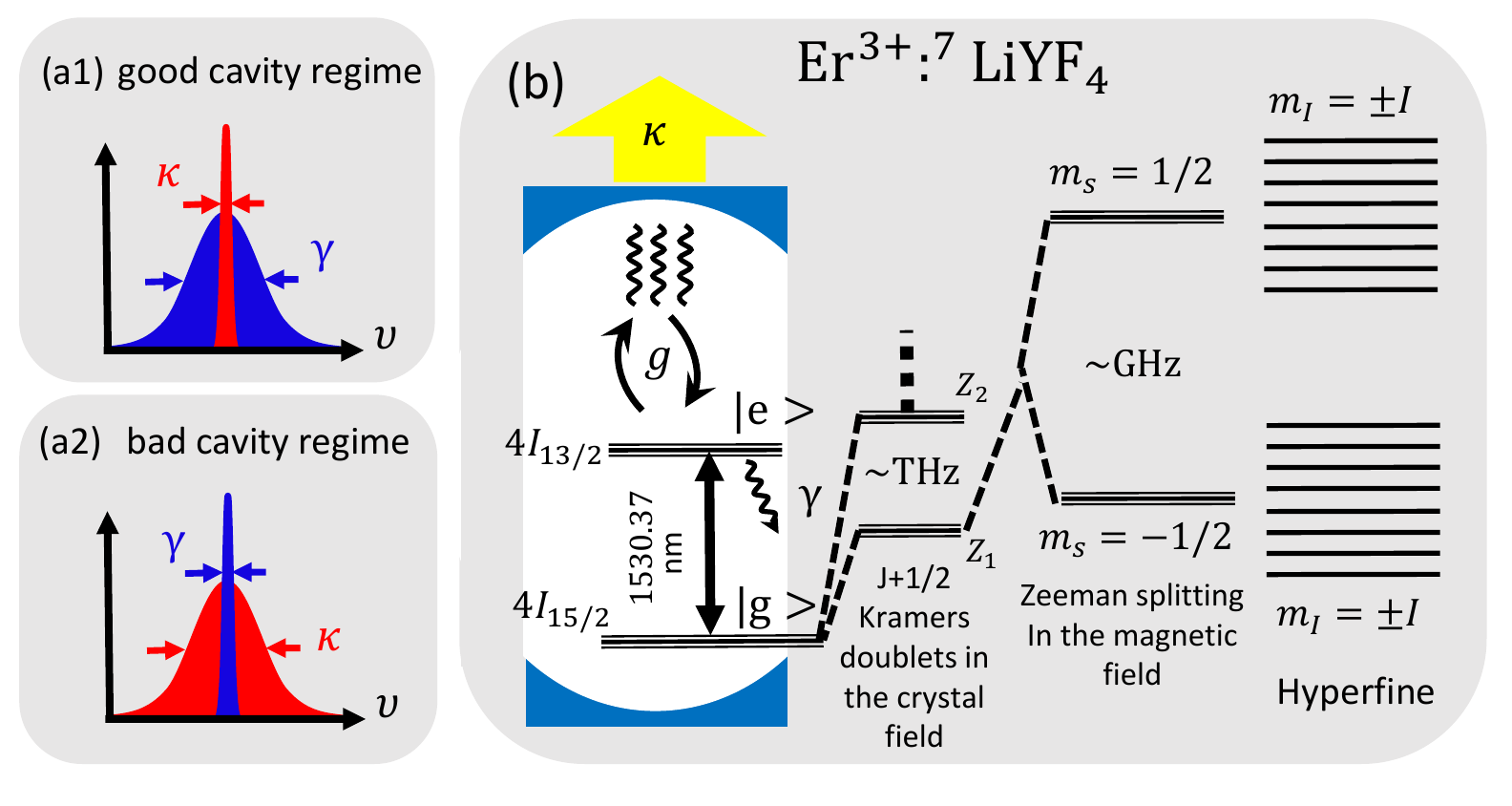}
    \caption{ (Color online) (a1(2)) The good (bad) cavity regime when the gain medium linewidth is much wider(narrower) than cavity linewidth. (b) The light-atom interaction in the cavity. $\kappa=-(c/2L)ln(R_1R_2)$ is the cold-cavity loss rate. $\gamma$ is the atomic decay rate and $g$ is the single-atom coupling to the cavity mode. A typical hyperfine structure of \erlyf\/ shown with the narrowest inhomogeneous broadening in RE ($\sim 16$ MHz) \cite{Thiel2011,Kukharchyk2018a,Gerasimov2016}. Even isotopes with nuclear spin $I=0$ is more interesting for a single mode SRL. At cryogenic temperature, only the lowest doublet $Z_1$ is populated, therefore the system can be described as an effective electronic spin with $S = 1/2$. More details discussed in the supplementary materials. }
    \label{Fig1}
\end{figure}

An alternative solution is to build up the coherence between atoms instead of photons. In this approach, atoms become spontaneously correlated, creating collective atomic dipole that emits light whose phase stability directly reflects the phase stability of the atomic dipole \cite{Meiser2009,Norcia2016a}. This phenomena, known as \textit{superradiance}, which the collective atomic dipole radiate a field whose intensity is proportional to the square of the number of atom while the radiative atomic decays inversely proportional to the number of atom \cite{Gross1982}. Actually, in a traditional laser, the gain medium linewidth is much wider than cavity linewidth (good-cavity regime, Fig. \ref{Fig1}a) while in the superradiant laser (SRL) regime the gain medium linewidth is much smaller than cavity linewidth (bad-cavity regime, Fig. \ref{Fig1}a) \cite{Kuppens1994}.

\par
Recently, a superradiant or bad-cavity laser is demonstrated on the millihertz linewidth strontium clock transition \cite{Norcia2016a}. The ultranarrow transitions is one of the main reasons that made alkaline-earth-metal-like atoms as a prime candidates for realizing such systems. The superradiant laser linewidth is proportionally related to the single-atom cooperativity times the atomic decay rate while larger number of atoms needed to enhance the collective phenomena through superradiance \cite{Debnath2018,Meiser2009}.

\begin{figure*}[t!]
$\begin{array}{rl}
    \includegraphics[width=0.5\textwidth]{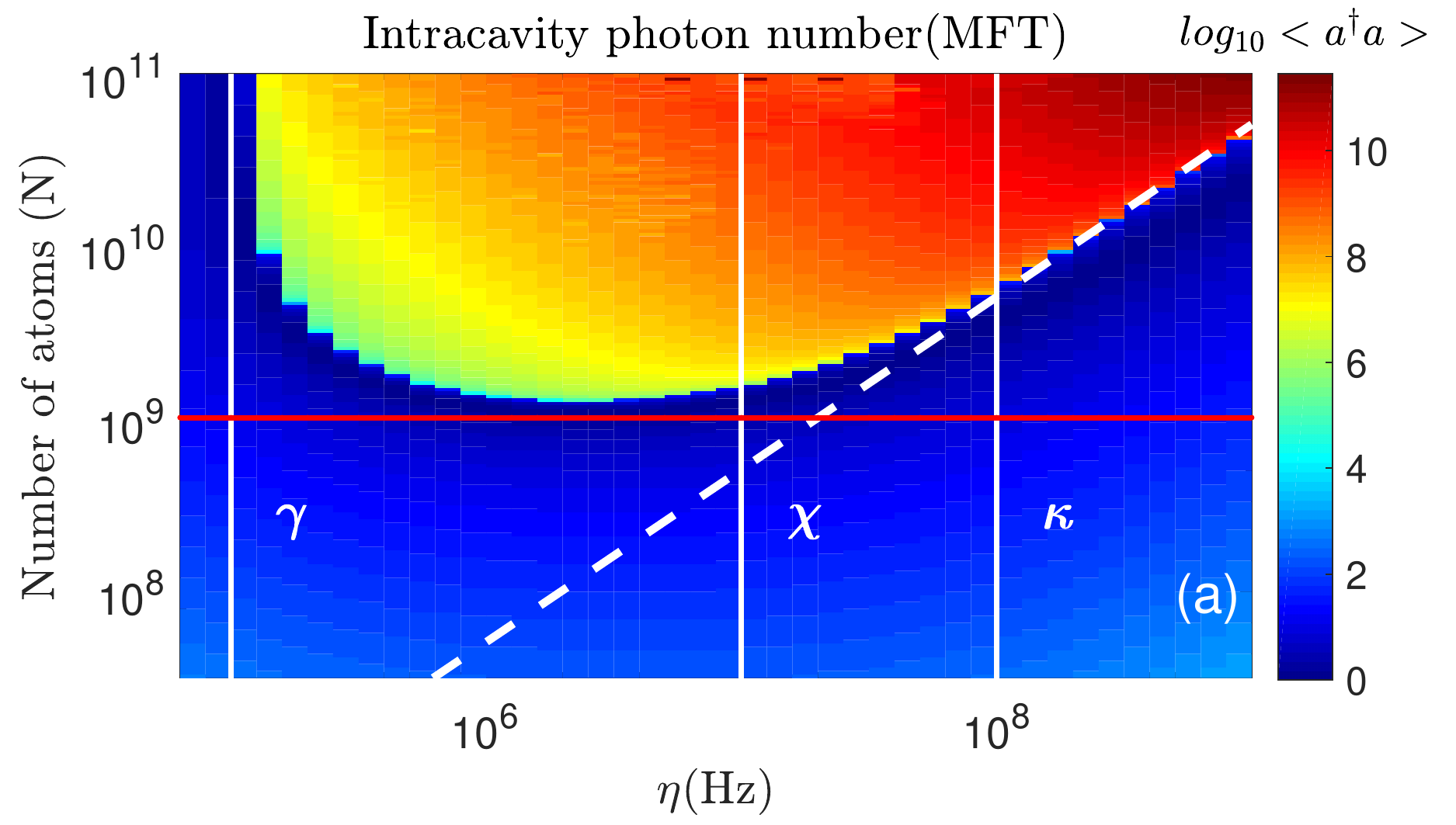} &
    \includegraphics[width=0.5\textwidth]{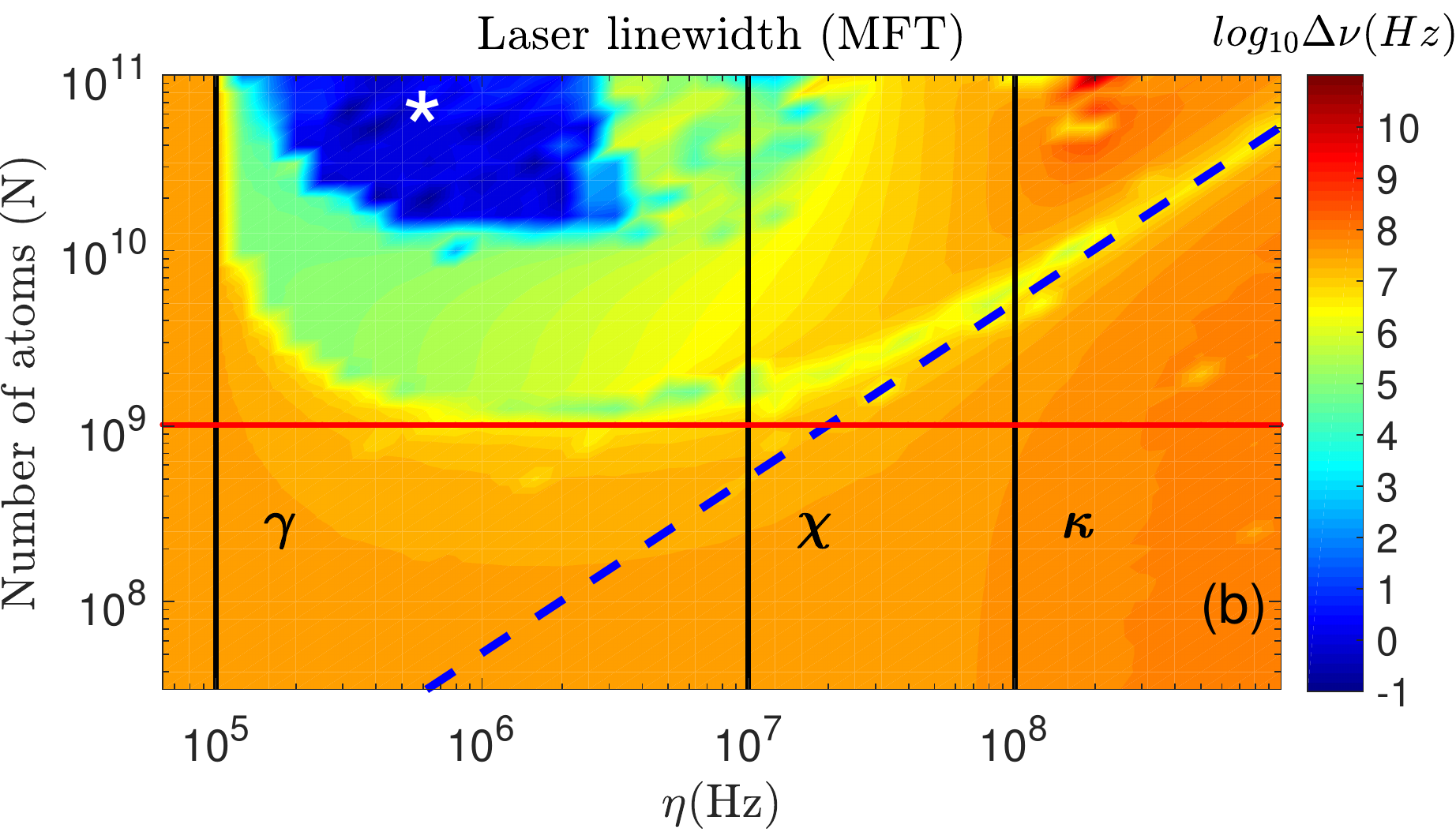}\\
\end{array}$
\caption {(Color online) For different atom number $N$ and incoherent pumping rate $\eta$ (a) The intracavity photon number and (b) The SRL linewidth is shown based on MFT calculation. $\gamma=100$kHz, $\chi= 10$ MHz, and $\kappa=100$ MHz are the atomic decay rate, inhomogeneous broadening, and cavity linewidth, respectively. The coupling constant $g=1.4$kHz. The white-dashed line is the maximum pumping rate $NC_1\gamma$. and red-solid line shows the critical atom number. \label{fig:Fig2}}
\end{figure*}

\par
In this letter, we take a step towards realization of narrow-band frequency standard in solid state materials. The target physical system proposed in this letter is based on naturally trapped rare-earth-ions (REIs) in a host crystal. The relatively long coherence time of REIs is the main attraction of these materials. In addition, higher atomic density and much smaller single atom-photon cooperativity compared to alkaline-earth-metal-like vapours makes REIs an interesting candidate for this purpose \cite{Thiel2011}. This proposal is supported by calculating the intracavity photon number and laser linewidth employing two different techniques. The master equation (ME) of the system is discussed and the numerical simulations based on \textit{Qutip} \cite{Shammah2018,Lambert2016,Johansson2012} shows a good agreement with the mean-field theory (MFT) approximation (supplementary materials). The master equation approach give us a quantum insights to the problem while the number of atoms ($\sim 50$) are limited to the computational power in the classical computers while the MFT approximation provides us the opportunity to calculate the intracavity photon number and the laser linewidth up to large number of atoms ($\sim 10^{10}$).

\par
%\textbf{Theoretical background:}
The theory of laser linewidth was formulated by Schawlow and Towns \cite{Schawlow1958} where the quantum limited linewidth for a homogeneously broadened single-mode laser tuned to the center of the gain profile given by:
   \begin{equation} {\label{eq:LLgoodcavity}}
    \Delta \nu= \frac{h\nu}{4\pi} \frac{\kappa^2}{P_{out}}
   \end{equation}
where $\Delta\nu$ is the (FWHM) laser linewidth, $\kappa=-(c/2L)ln(R_1R_2)$ is the cold-cavity loss rate (see Fig. \ref{Fig1}a), with $L$ the cavity length and $R_1$ and $R_2$ the mirror reflectivities, and $P_{out}$ is the laser output power. Eq. \ref{eq:LLgoodcavity} has been derived under assumption that the gain bandwidth (FWHM), denoted by $2\gamma = 2/T_2$ is much larger than the cavity loss rate $\kappa$, i.e., $a \equiv \kappa/2\gamma \ll 1$ which is called the good-cavity limit.

\par
The homogeneously broadened single-mode laser in the bad-cavity regime is described in Ref.\cite{Haken1984} as follows:
   \begin{equation} {\label{eq:LLbadCavity}}
    \Delta \nu= \frac{h\nu}{4\pi} \frac{(\kappa /n_g)^2}{P_{out}}N_{sp}(1+[\frac{2\pi(\nu-\nu_0)}{\gamma+\frac{1}{2}\kappa}]^2)
   \end{equation}
where the spontaneous emission factor $N_{sp}=N_e/(N_e-N_g)$ measure the degree of inversion where $N_e$ and $N_g$ are excited and ground state population, respectively. Assuming zero detuning ($\nu \sim \nu_0$), the main factor which affect the laser linewidth is the group refractive index $n_g=(\frac{2\gamma+\kappa}{2\gamma})$. This reflects the memory effect of the polarization that effectively slow down the phase diffusion process \cite{Kolobov1993}.
A system including a cavity and an atomic transition, oscillates at frequency $f=(2\gamma f_{cavity}+\kappa f_{atomic})/(2\gamma+\kappa)$. The sensitivity to change of the system frequency with respect to a change in cavity frequency, is called frequency pulling coefficient $P=\frac{df}{df_{cav}}=\frac{2\gamma}{2\gamma+\kappa}=\frac{1}{n_g}$. In the good-cavity case, where $a\equiv \kappa/2\gamma \ll1$, the frequency pulling coefficient will be $1$ while in the bad-cavity case, $a\equiv \kappa/2\gamma \gg1$, we will have $P=2\gamma/\kappa \ll1$. Therefore, to decrease the laser linewidth, one needs to reduce the frequency pulling coefficient or increase the group refractive index.

\par
Indeed, the group refractive index $n_g(\nu)=n_r+\nu \frac{dn}{d\nu}$ can be significantly deviate from the real refractive index $n_r \approx 1$ due to steep gradient of the atomic transition. As discussed in ref. \cite{Kuppens1994} one can replace the cold-cavity loss rate $\kappa$ by a dressed loss rate $\kappa/n_g$. As shown in ref. \cite{Sabooni2013b}, one can obtain at least four order of magnitude larger group refractive index compared to the real refractive index in the rare-earth ion doped crystals.

\par
Following Eq. \ref{eq:LLbadCavity}, the laser linewidth for a good cavity ($a\equiv \kappa/2\gamma \ll1$) at resonance frequency will be $\Delta \nu_{GC}= \frac{h\nu}{4\pi} \frac{\kappa^2}{P_{out}}$. By replacing $P_{out}=E.\kappa=M_c.h\nu.\kappa$, the good-cavity laser linewidth could be written in terms of intracavity photon number, $\Delta \nu_{GC}= \frac{\kappa}{4\pi} \frac{1}{M_c}$, where $M_c$ is the average intracavity photon number. In the bad-cavity case ($a\equiv \kappa/2\gamma \gg1$), the group refractive index will be $n_g \approx \frac{\kappa}{2\gamma}$, therefore the laser linewidth expression will be:
   \begin{equation} {\label{eq:LLgoodcavity2}}
        \Delta \nu_{BC}= \frac{\gamma^2}{\kappa \pi} \frac{1}{M_c}
   \end{equation}

\par
The ultimate goal is to place the REI crystals inside a cavity and increase the coupling interaction between light and atoms in a collective manner. As shown in Fig. \ref{Fig1}b, the dynamics of the system could be described by three rate: the coupling between single atom and single photon ($g$), the cavity decay ($\kappa$), the atom spontaneous emission rate of a two level transition ($\gamma$). Important parameters for atom-cavity characterization are the two dimensionless parameters called the critical atom number ($N_c=\frac{\gamma \kappa}{g^2}$) and the saturation photon number ($M_c=\frac{\gamma^2}{g^2}$) \cite{Grimm2000}. Therefore the relation between critical atom number and saturation photon number will be $\frac{N_c}{M_c}=\frac{\kappa}{\gamma}$. The cooperativity has a inverse relation with the critical atom number $N_c \propto \frac{1}{C}$ therefore, one can rewrite the Eq. \ref{eq:LLgoodcavity2} in terms of the single atom cooperativity as follows:
   \begin{equation} {\label{eq:LLbadcavLine_1}}
        \Delta \nu_{BC}= \frac{C_1\gamma}{\pi}
   \end{equation}
where $C_1=\frac{g^2}{\gamma\kappa}$ is the single atom cooperativity and $g$ is the single-atom coupling to the cavity emission mode. Therefore, to decrease the laser linewidth, one needs to have smaller as possible single atom cooperativity while needs large number of atom to be collectively enhanced through the pumping rate of $NC_1\gamma$. One can write the single atom cooperativity as $C_1=\frac{\mathcal{F}\sigma_0}{A}$, where $\mathcal{F}$ is the cavity finesse, $\sigma_0$ is the resonant absorption cross section \cite{Pernas2005,Hilborn1982}, and $A$ is the effective beam area. Therefore, the laser linewidth will be: $\Delta \nu_{BC}= \frac{c}{2\pi n_g L} \frac{\sigma_0}{A} \frac{\gamma}{\kappa}$ where $L$ is the cavity length and $c$ is the speed of light.

\begin{figure*}[t!]
$\begin{array}{rl}
    \includegraphics[width=0.4\textwidth]{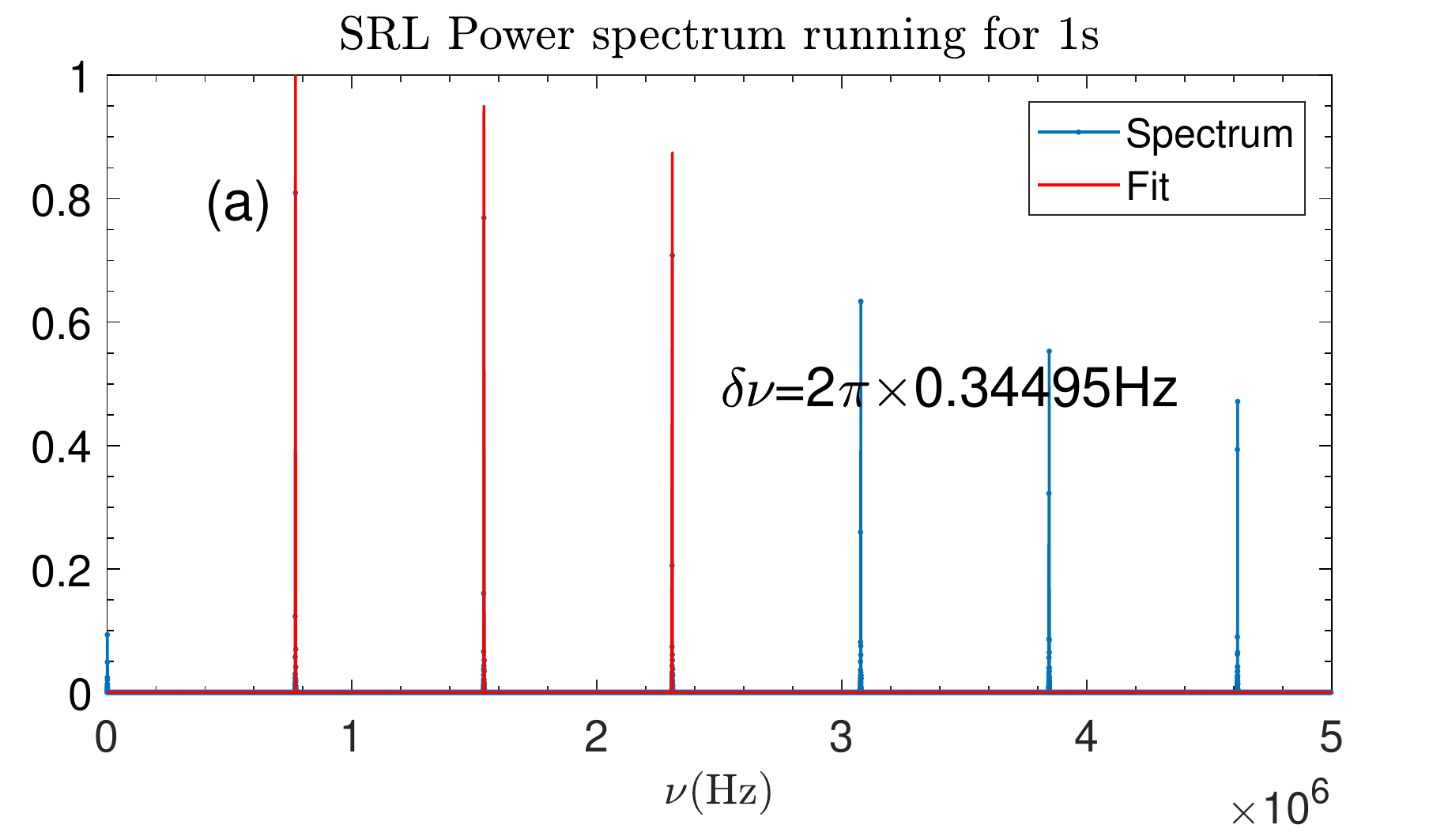} &
    \includegraphics[width=0.4\textwidth]{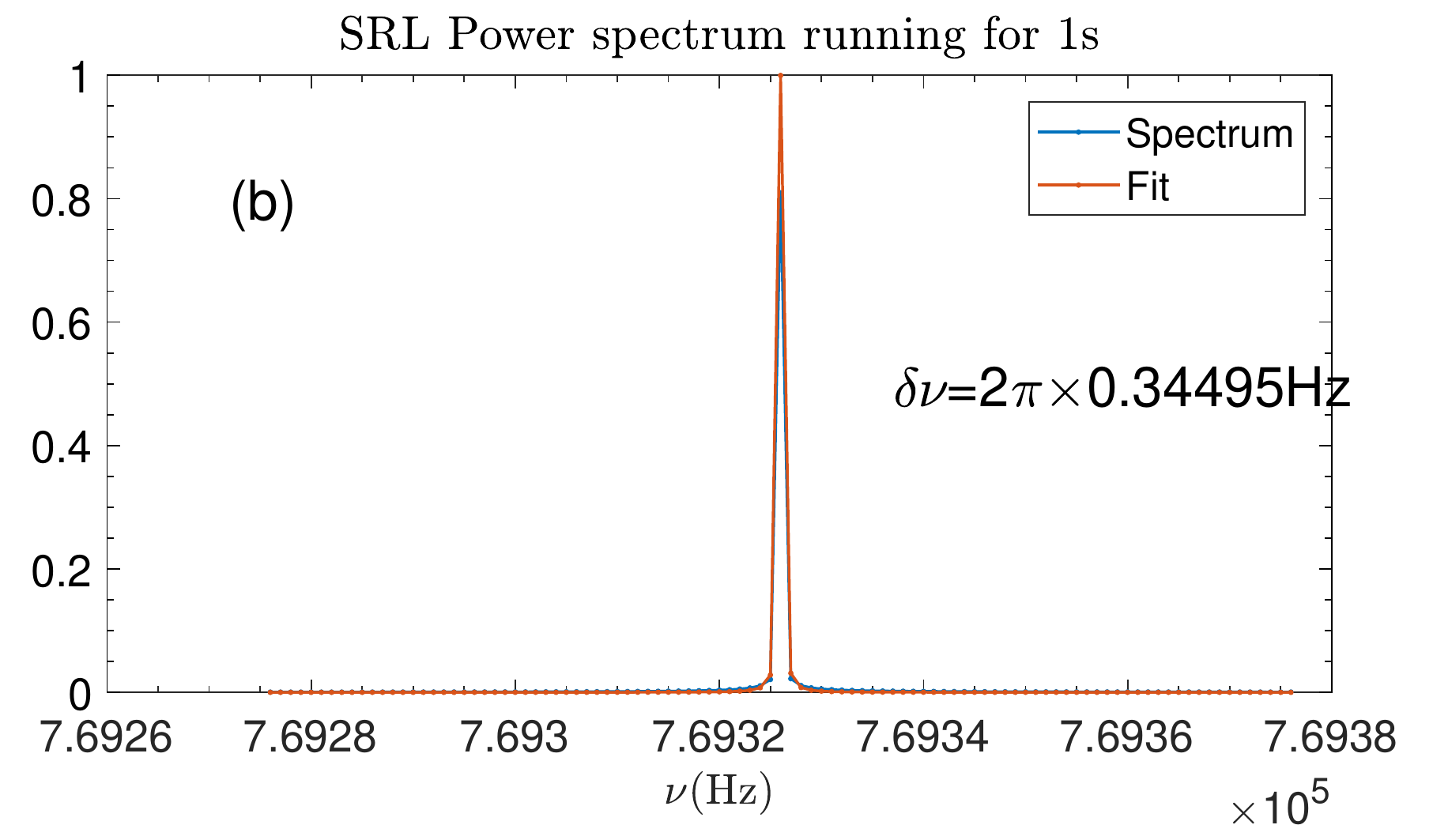}\\
\end{array}$
\caption {(Color online) A typical narrow linewidth SRL spectrum. (a) The full spectrum (b) The calculated linewidth for a single mode SRL emission. The sampling rate is $100$ MHz and the SRL steady-state emission duration is $1$s. This is corresponding power spectrum of the white asterisk in Fig. \ref{fig:Fig2}b.\label{fig:SPEC}}
\end{figure*}

\par
%\textbf{Simulation:}
The photon emitted to the cavity mode mediated atoms to be correlated \cite{Gross1982,Dicke1954}. This will lock the phase of the atomic dipoles to act as a macroscopic dipole. Eventually, this macroscopic dipole will be less sensitive to the environmental noise and has potential to reduced the output light emitted linewidth. The collective interaction of closed packed spins in superradiance makes mean-field theory an intuitively the best first attempt solution. The intracavity mean photon number equation rate could be related to the atom-field correlation as follows:

\begin{equation} {\label{eq:MFTEq1}}
\begin{split}
\frac{d\langle \hat{a}^{\dagger} \hat{a}\rangle}{dt} &=  -\kappa \langle \hat{a}^{\dagger} \hat{a}\rangle
 -\frac{igN}{2}[\langle \hat{a}^{\dagger} \hat{\sigma}^-_1\rangle - \langle \hat{\sigma}^+_1\hat{a} \rangle]\\
\end{split}
\end{equation}
Where $\kappa$ is the cavity decay, $g$ is the single atom-photon coupling strength, and $N$ represent the number of atoms. The atom-field $\langle \hat{a}^{\dagger} \hat{\sigma}^-_1\rangle$ coherence evolves according to:

\begin{equation} {\label{eq:MFTEq2}}
\begin{split}
\frac{d\langle \hat{a}^{\dagger} \hat{\sigma}^-_1\rangle}{dt}=  -(\frac{\eta+\gamma+\kappa}{2}+\chi-i\delta)\langle \hat{a}^{\dagger} \hat{\sigma}^-_1\rangle \\
      + \frac{ig}{2}[ \langle \hat{\sigma}^z\rangle \langle \hat{a}^{\dagger} \hat{a}\rangle +\frac{\langle \hat{\sigma}^z\rangle+1}{2}+(N-1)\langle \hat{\sigma}^+_1 \hat{\sigma}^-_2\rangle  ] \\
\end{split}
\end{equation}
where $\delta=\omega_c-\omega_a$ is the cavity-atom detuning frequency. The steady-state of our system considered to run at $\delta=0$. The $\eta$ and $\gamma$ are incoherent pumping and atomic decay, respectively. $\chi=\frac{1}{T^*_2}$ is the representation of the inhomogeneously broadened sample. The higher order correlations are ignored in all equations. The rate of population inversion through the atom-field coupling will be as follows:

\begin{equation} {\label{eq:MFTEq3}}
\begin{split}
\frac{d\langle \hat{\sigma}^z\rangle}{dt}=ig(\langle \hat{a}^{\dagger} \hat{\sigma}^-_1\rangle -\langle \hat{\sigma}^+_1\hat{a}\rangle)
-\gamma(1+\langle \hat{\sigma}^z\rangle) +\eta(1-\langle \hat{\sigma}^z\rangle)         \\
\end{split}
\end{equation}
and finally, to close the sets of equations, the spin-spin correlations evolve according to:
\begin{equation} {\label{eq:MFTEq4}}
\begin{split}
\frac{d\langle \hat{\sigma}^+_1 \hat{\sigma}^-_2\rangle}{dt}=  -(\gamma+\eta+2\chi)\langle \hat{\sigma}^+_1 \hat{\sigma}^-_2\rangle\\
-\frac{i g}{2}\langle \hat{\sigma}^z\rangle[\langle \hat{a}^{\dagger} \hat{\sigma}^-_1\rangle-\langle \hat{\sigma}^+_1 \hat{a}\rangle]\\
\end{split}
\end{equation}
\par
It has been verified that the steady-state is much faster than the anticipated total operation time \cite{Meiser2009}, therefore, we will calculate the steady-state intracavity photon number employing Eqs. \ref{eq:MFTEq1}-\ref{eq:MFTEq4} in a set of four ordinary differential equations (ODEs), while we have set all four initial conditions to be zero. The exact solution for steady-state intracavity photon number versus incoherent pumping rate $\eta$ and atom number $N$ shown in Fig.\ref{fig:Fig2}a.

\par
Obtaining the spectrum of the SRL emitted photon is a vital step towards proving the strength of SRL as a ultra-narrow optical frequency linewidth photon source. The quantum regression theorem is employed to find the equations of motion for the first-order two time correlation function of the light field $\langle \hat{a}^\dagger(t)\hat{a}(0)\rangle$ \cite{Carmichael1999}. This will be nothing more than solving set of four closed ODEs (Eqs.\ref{eq:MFTEq1}-\ref{eq:MFTEq4}) while employing steady-state solution achieved in the first place as an initial conditions. Given the first-order correlation function $\langle \hat{a}^\dagger(t)\hat{a}(0)\rangle$, one can define the corresponding power spectrum as follows: $S(\omega)=\int_{-\infty}^{+\infty} \langle \hat{a}^\dagger(t)\hat{a}(0)\rangle e^{-i\omega t} dt$.
The SRL linewidth contour plot for different atom number $N$ versus incoherent pumping rate $\eta$ is shown in Fig.\ref{fig:Fig2}b. A typical SRL spectrum is shown in Fig. \ref{fig:SPEC}a and with more details in Fig.\ref{fig:SPEC}b. This is corresponding power spectrum of the white asterisk in Fig. \ref{fig:Fig2}b. This extremely narrow SRL linewidth is predicted to happen in a area where $\gamma<\eta<\kappa$ and $n>n_{crit}$ as shown in Fig.\ref{fig:Fig2}. A Lorentzian function $F(\nu;\nu_0,A,\sigma)=\frac{A}{\pi}[\frac{\sigma}{(\nu-\nu_0)^2+\sigma^2}]$ is employed to fit the SRL spectrum where the parameter amplitude corresponds to $A$, center to $\nu_0$, and half-width-half-maximum to $\sigma$.

\par
There are three main criteria for having narrow linewidth steady-state-superradiance-radiation (SSSR). Firstly, the atom number should be higher than the critical atom number $N_{crit}=\frac{2\chi}{C_1\gamma}$ . Secondly, the collective decay much larger than other source of decay and decoherence rate ($NC_1 \gamma \gg \gamma,\chi $) and thirdly, the pumping rate $\eta$ smaller than the collective decay rate ($w \approx NC_1\gamma$) to produce a high rate of intracavity photon while $\gamma<\eta<\chi$ is a necessary condition to have SRL linewidth smaller than the atomic decay rate $\gamma$ . The quantity $NC_1$ has no relationship with the excitation mode volume but in \erlyf \/ is high enough to satisfy all conditions for having SSSR mentioned above.

\par
Isotopically purified RE-doped LYF crystals are well known for their ultra-narrow optical inhomogeneous broadening which is limited by super-hyperfine interactions between electronic spins of impurity ions and nuclei spins of the host crystal \cite{Thiel2011}. The clock transition of even isotopes with nuclear spin $I=0$ could be a suitable candidate for SRL emission \cite{Kukharchyk2018a}.
\erlyf\/ has the narrowest inhomogeneously broadened transition in REI solids ($\Gamma_{inh}\sim 16$ MHz) \cite{Thiel2011,Kukharchyk2018a,Gerasimov2016}. In \eryso sample a $\Gamma_{inh}\sim 12$ MHz is reported \cite{Probst2013}. As shown in Fig. \ref{Fig1s} the \dhdhhh transition in \erlyf\/ at $1530.372$ nm with $T_1=9.5$ ms and $T_2 \sim 100 \mu$s and $\mu_{eg}= 2.72 \times 10^{-32}$ C.m is suggested. The optical clock transition for one of Zeeman transition (transition 3) at $B=0.2T$ is shown in Fig. 2b of Ref. \cite{Kukharchyk2018a}. The single atom cooperativity in \erlyf\/ at $1.53\mu$m for $100 \mu$m beam radius and $1$mm cavity length could be as low as $C_1=10^{-8}$ \cite{Leuchs2013}.

\par
The number of ions $N_i$ within the homogeneously broadened frequency channel, on average, within this excitation volume $V_{ex}$ is approximated as $N_i=D_h.C_d.\frac{\Gamma_h}{\Gamma_{inh}}.V_{ex}$ where $D_h$ is the density of the host ions which is in the order of several $10^{10}\mu m^{-3}$ \cite{Moksimov1971}. The doping concentration could be arrange from several percent to very dilute ($\sim 1-0.1$ppm). The narrow transition ($\Gamma_{inh}=16$ MHz) of \erlyf measured at $C_d \approx 0.005\%$. $\Gamma_h$ and $\Gamma_{inh}$ are homogeneous and inhomogeneous broadening of the absorption cross section. The number of atom per excitation mode volume for \erlyf \/ could be up to $\sim 10^{11}$. The total cooperativity $NC_1 \sim 10^3$. Stoichiometric REI crystals with smaller inhomogeneous broadening and at least several order of magnitude more optical depth are potentially good candidates for SRL \cite{Ahlefeldt2016}. One needs to measure the SRL linewidth as an important benchmark to be compared with the ordinary lasers. One of the main experimental methods for measuring the laser linewidth is based on heterodyne beat note created via two stable lasers. In our proposed method one can demonstrate two superradiant laser in just a single crystal but in a different spatial positions.

\par
%\textbf{Clock transition stability:}
REI based SRL is a promising candidate as a laser local oscillator if one could find a clock transition insensitive against environmental noise. As a benchmark, the level of accuracy needed to measure the gravitation redshift for distance of $\Delta h=10$ cm for center frequency of $\nu_0\sim 200$ THz is $\delta \nu/\nu_0=g\Delta h/c^2 = 10^{-17}$. Therefore, we needs to have $\delta \nu\sim 2$ mHz. The collective SRL emission act as active phase locking system between individual ions and can fill the gap between the atomic homogeneous linewidth and required frequency linewidth $\delta\nu$ for quantum metrology through small single atom cooperativity $C_1$. The temperature sensitivity of the spectral-hole frequency was measured in a typical REI crystal in Ref. \cite{Thorpe2011} to be about $16kHz/K^{-2}$. This means a typical REI with $\Gamma_h\sim1$kHz needs to be stable with sensitivity of about $0.25$ K at target operation temperature of smaller than $2$ K \cite{Lovis2006}. In addition, the pressure sensitivity was reported to be about $211.4 \frac{Hz}{Pa}$ \cite{Thorpe2011} which means we need to control the pressure with $\sim 5$Pa accuracy. The crystal acceleration measured  up to $7\times 10^{-12} g^{-1} (1g=9.8ms^{-2})$ in Ref. \cite{Thorpe2011} which is corresponding to the frequency shift of $\sim0.25$ Hz and well below the passive acceleration-sensitivity of FP cavities and above $\Gamma_h\sim1$kHz \cite{Jiang2011,Webster2007,Konz2003}. The most important environmental perturbations are magnetic and electric field noise. The curvature of the transition frequency with respect to the magnetic field is reported about $\sim 130 \frac{Hz}{G^2}$ at the field of $\sim 0.28$ T \cite{Kukharchyk2018a}. Therefore, one needs to control the environmental magnetic field with accuracy of $\sim 3$ G. The linear Stark effect on the \dhdhhh transition in an Erbium is reported to be $10\/ \frac{kHz}{Vcm^{-1}}$ \cite{Hastings-Simon2006}, therefore, having controllability of $300\/ \frac{mV}{cm^{-1}}$ over the sample is necessary.

%\section{Conclusion\label{sec:conclusion}}
\par
As a conclusion, an active solid-state optical clock superradiant laser, which is almost insensitive to the cavity mirror vibration is proposed. The proposed narrow linewidth REI superradiant laser (RESALE) can compete with the best frequency-stable local oscillators. The long coherence time and the large optical density of Rare-Earth-ions (REI) doped solids are employed to find a regime to demonstrate a steady-state laser emission with a linewidth smaller than the atomic decay rate. The combination of the small single-atom cooperativity, the large optical density, and the long coherence time of REI provide the possibility to realize a steady-state sub-Hz level laser emission. The RESALE proposal has a great potential to extend the application of REI to the quantum metrology.

%\section{Acknowledgement\label{sec:Acknow}}
\par
I thank K. S. Choi, S. A. Moiseev, M. N. Popova, K. I. Gerasimov, S. Kr\"{o}ll, and P. Bushev for the stimulating discussion of this work.

%\bibliography{C:/Users/Mahmood/Dropbox/PAPERS/Main_Ref}
%\bibliography{E:/Info-document/files/Polarization_rotation/Article/Main_Ref}
%\input{CavityQMNJP.bbl}

%%%%%%%%%%%%%%%%%%%%%%%%%%%%%%%%%%%%%%%%%%%%%%%%%%%%%%%%%%%%%%%%%%%%%%%%%%%%%%%%%%%%%%%%%%%%%%%%%%%%%%%%%%%%%%%%%%%%%%%%%%%%%%%%%%%%%%%%%%%%%%%%%%%%%%%%%
%%%%%%%%%%%%%%%%%%%%%%%%%%%%%%%%%%%%%%%%%%%%%%%%%%%%%%%%%%%%%%%%%%%%%%%%%%%%%%%%%%%%%%%%%%%%%%%%%%%%%%%%%%%%%%%%%%%%%%%%%%%%%%%%%%%%%%%%%%%%%%%%%%%%%%%%%
%%%%%%%%%%%%%%%%%%%%%%%%%%%%%%%%%%%%%%%%%%%%%%%%%%%%%%%%%%%%%%%%%%%%%%%%%%%%%%%%%%%%%%%%%%%%%%%%%%%%%%%%%%%%%%%%%%%%%%%%%%%%%%%%%%%%%%%%%%%%%%%%%%%%%%%%%
%%%%%%%%%%%%%%%%%%%%%%%%%%%%%%%%%%%%%%%%%%%%%%%%%%%%%%%%%%%%%%%%%%%%%%%%%%%%%%%%%%%%%%%%%%%%%%%%%%%%%%%%%%%%%%%%%%%%%%%%%%%%%%%%%%%%%%%%%%%%%%%%%%%%%%%%%
%%%%%%%%%%%%%%%%%%%%%%%%%%%%%%%%%%%%%%%%%%%%%%%%%%%%%%%%%%%%%%%%%%%%%%%%%%%%%%%%%%%%%%%%%%%%%%%%%%%%%%%%%%%%%%%%%%%%%%%%%%%%%%%%%%%%%%%%%%%%%%%%%%%%%%%%%

%%%%%%%%%% Merge with supplemental materials %%%%%%%%%%
\pagebreak
%\widetext
\onecolumngrid
\clearpage
\begin{center}
\textbf{\large Supplemental Materials:}
\end{center}
%%%%%%%%%% Merge with supplemental materials %%%%%%%%%%
%%%%%%%%%% Prefix a "S" to all equations, figures, tables and reset the counter %%%%%%%%%%
\setcounter{equation}{0}
\setcounter{figure}{0}
\setcounter{table}{0}
\setcounter{page}{1}
\makeatletter
\renewcommand{\theequation}{S\arabic{equation}}
\renewcommand{\thefigure}{S\arabic{figure}}
%\renewcommand{\bibnumfmt}[1]{[S#1]}
%\renewcommand{\citenumfont}[1]{S#1}
%%%%%%%%%% Prefix a "S" to all equations, figures, tables and reset the counter %%%%%%%%%%
%\section{Section 1}
%Copy and paste your Supplemental Materials text here \cite{S_RefA}, blah, blah, blah, blah, blah, blah, ...
%\begin{equation}
%  i\hbar\frac{\partial}{\partial t}\psi(x,t) = -\frac{\hbar^2}{2m}\frac{\partial^2}{\partial x^2}\psi(x,t) + V(x,t) \psi(x,t)
%\end{equation}
\subsection{Master equation (ME) based solution:\label{subsec:ME}}
\begin{figure*}[ht]
$\begin{array}{rl}
    \includegraphics[width=0.5\textwidth]{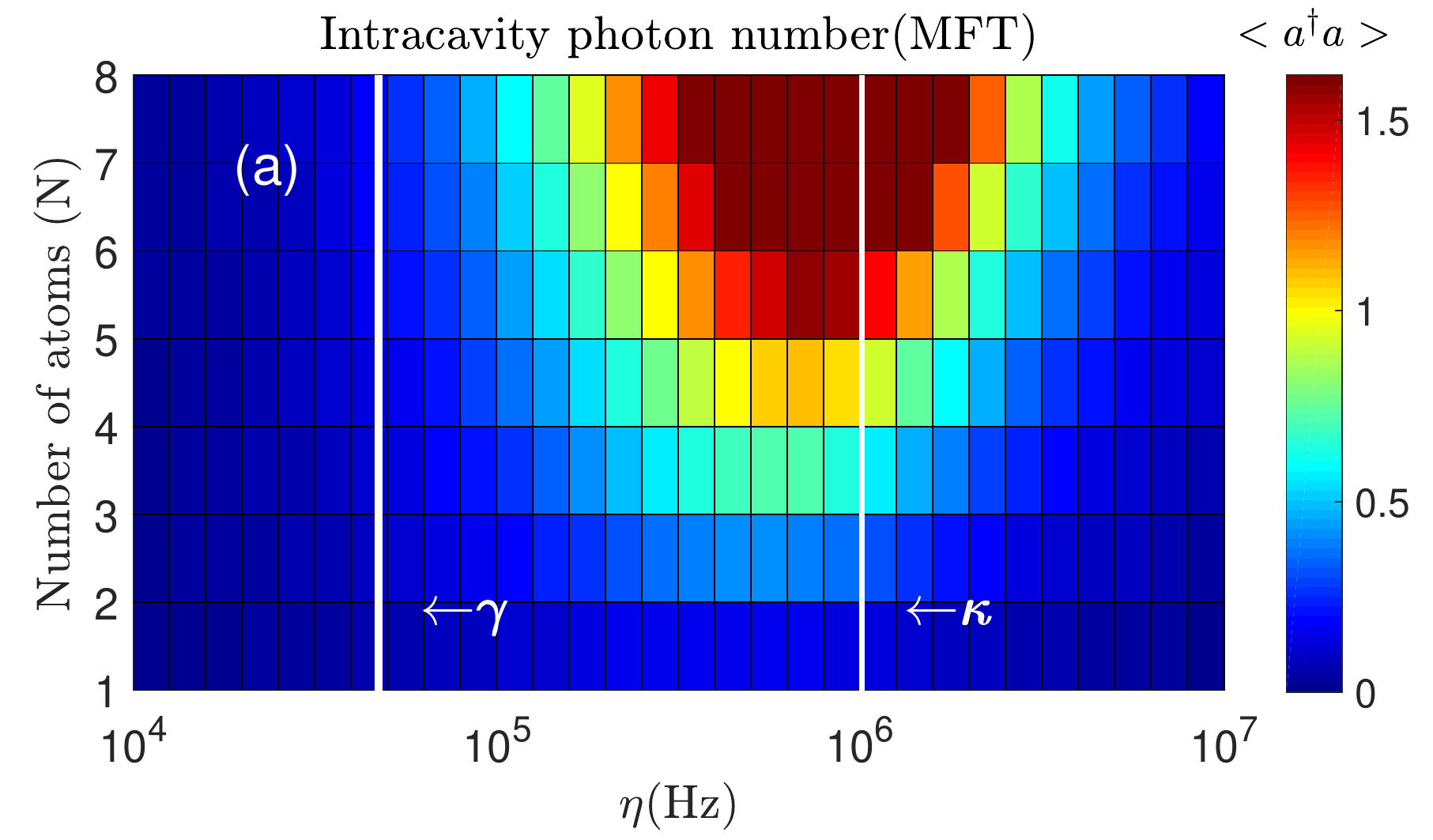} &
    \includegraphics[width=0.5\textwidth]{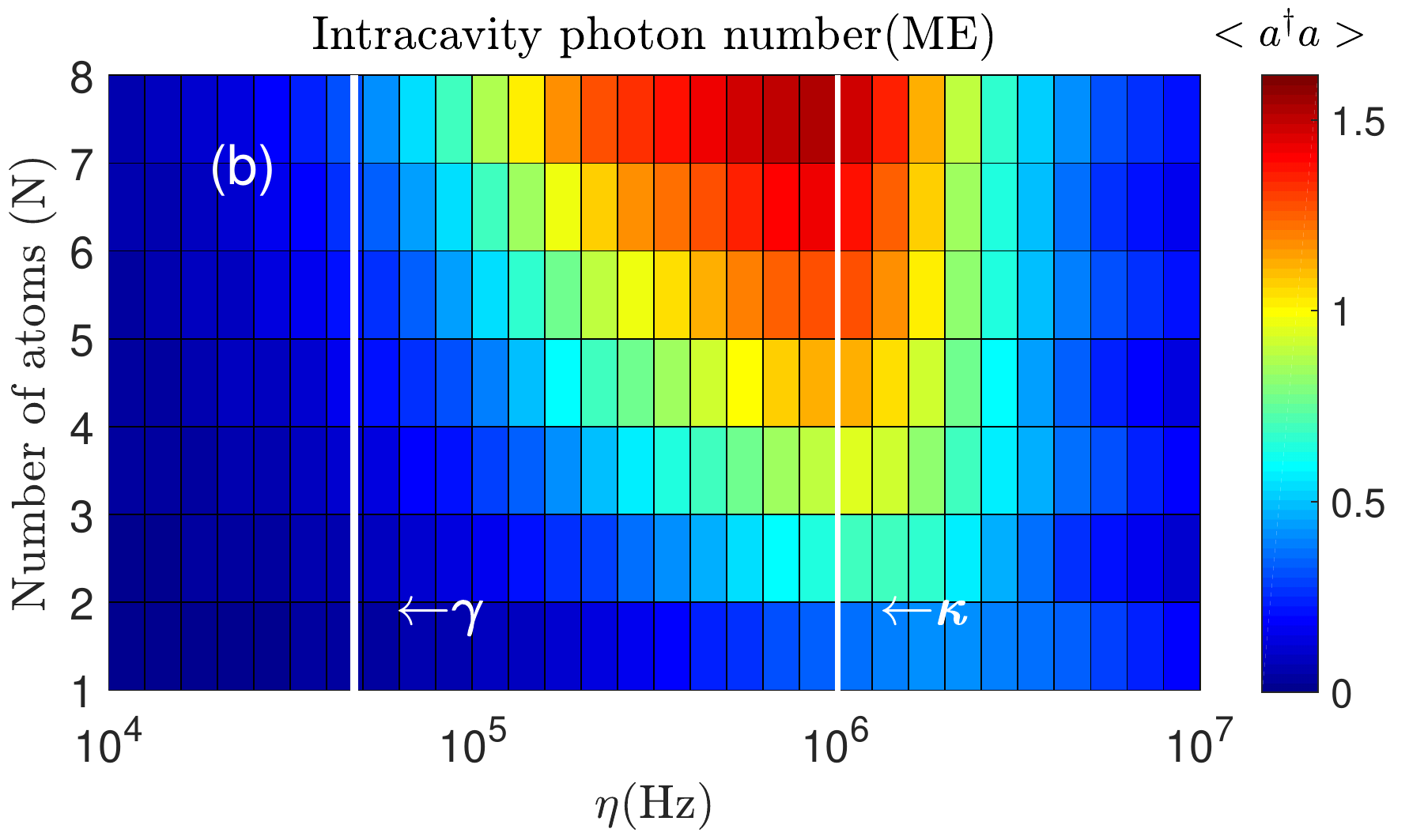}\\
\end{array}$
\caption {(Color online) The intracavity photon number $\langle a^{\dagger} a\rangle$ as a function of the local pumping rate $\eta$ and atom number $N$ based on (a) mean-field theory (Eqs. \ref{eq:MFTEq1}-\ref{eq:MFTEq4}) (b) master equation (Eq. \ref{eq:masterEqCase1}). The atomic decay rate $\gamma=2\pi \times 7.5$ kHz and the cavity linewidth $\kappa= 2\pi \times 160$ kHz is shown by white line. The coupling strength is $g= 2\pi \times 100$ kHz.\label{fig:FigMEMFT}}
\end{figure*}
\par
The dynamics of an open quantum system consisting of an ensemble of identical qubits that can dissipate through local and collective baths could be analyzed according to a Lindblad master equation. The Liouvillian of an ensemble of N qubits, or two-level systems (TLSs), can be built and solved using the Permutational Invariant Quantum Solver (PIQS) as a QuTiP module \cite{Shammah2018,Lambert2016,Johansson2012}. The master equation (ME) based solution provide us more deep understanding of entanglement properties of an open TLS system while it is limited by computational source to the low number of atoms ($\sim 100$). One of the main strength of REI based SRL simulation is the rather high atomic density, therefore, the ME will not be an optimum solution. However, the results of ME calculation is rather consistent with MFT simulation.
\par
A general two-level open quantum system (TLS) with local and collective interaction with a bosonic cavity through a coherent dynamics is identified as follows \cite{Shammah2018}:
\begin{equation} {\label{eq:masterEq}}
%\begin{gather*}
%\begin{split}
\dot{\rho}=- \frac{i}{\hbar}[H,\rho]\\
           +\left\{\frac{\gamma_{\Downarrow}}{2}\Lb_{J_-}[\rho]+\frac{\gamma_{\Uparrow}}{2}\Lb_{J_+}[\rho]+\frac{\gamma_{\Phi}}{2}\Lb_{J_z}[\rho]\right\}\\
           +\left\{\sum\limits_{n=1}^N(\frac{\gamma_{\downarrow}}{2}\Lb_{J_{-,n}}[\rho]+\frac{\gamma_{\uparrow}}{2}\Lb_{J_{+,n}}[\rho]+\frac{\gamma_{\phi}}{2}\Lb_{J_{z-,n}}[\rho])\right\}
%\end{split}
%\end{gather*}
\end{equation}

where $\rho$ is the density matrix of the full system and $H$ is the TLS ensemble Hamiltonian. Here $[J_{x,n},J_{y,m}]=i\delta_{m,n}J_{z,n}$,$[J_{+,n},J_{-,m}]=2\delta_{m,n}J_{z,n}$, and $J_{\pm,n}=J_{x,n}\pm iJ_{y,n}$. The spin operators $J_{\alpha,n}=\frac{1}{2}\sigma_{\alpha,n}$ for $\alpha=\left\{x,y,z\right\}$ and $J_{\pm,n}=\sigma_{\pm,n}$. The \textit{Lindblad} superoperators defined by $\Lb_A[\rho]=2A\rho A^{\dagger}-A^{\dagger} A \rho -\rho A^{\dagger} A$ and $\gamma_{\Downarrow},\gamma_{\Uparrow},\gamma_{\Phi},\gamma_{\downarrow},\gamma_{\uparrow}$, and$\gamma_{\phi}$ are the coefficients characterizing collective emission, collective pumping, collective dephasing, homogeneous local emission (radiative and non-radiative losses, $\gamma_{\downarrow}=\gamma_0(1-n_T)$), homogeneous local pumping ($\gamma_{\uparrow}=\gamma_0 n_T$), and homogeneous local dephasing, respectively. $\gamma_0$ is fixed for given system and $n_T$ is the thermal population of environment. Eq. \ref{eq:masterEq} is derived under Markov approximation( environment memory-less), Born Approximation( system and environment always stays in product state), and RWA approximation ($\gamma_i$ coefficients of Eq. \ref{eq:masterEq} is much smaller than the coupling present in the Hamiltonian). For the case of SRL in REI, the master equation \ref{eq:masterEq} is simplified by ignoring $\gamma_{\Phi},\gamma_{\phi},$ and $\gamma_{\Uparrow}$. Therefore, the master equation will be:

\begin{equation} {\label{eq:masterEqCase1}}
%\begin{split}
\dot{\rho}=-\frac{i}{\hbar}[H,\rho]+\frac{\gamma_{\Downarrow}}{2}\Lb_{J_-}[\rho]\\ +\sum\limits_{n=1}^N(\frac{\gamma_{\downarrow}}{2}\Lb_{J_{-,n}}[\rho]+\frac{\gamma_{\uparrow}}{2}\Lb_{J_{+,n}}[\rho])
%\end{split}
\end{equation}
The \textit{Qutip} numerical simulation package is employed to solve Eq. \ref{eq:masterEqCase1}. The intracavity photon number for the steady-state $\langle a^{\dagger} a\rangle$, as a function of the local pumping rate $\gamma_{\uparrow}=\eta$ and atom number N is shown in Fig. \ref{fig:FigMEMFT}a. The atomic decay rate $\gamma=7.5$ kHz and the cavity linewidth $\kappa= 2\pi \times 160$ kHz are shown in white-solid lines. Comparing Fig. \ref{fig:FigMEMFT}a and Fig. \ref{fig:FigMEMFT}b gives us an insight about the consistency of ME and MFT solutions at least in low atom number limit.

\subsection{Spectrum of $^{166}$\erlyf \label{sec:erlyf}}
\begin{figure}[ht]
    \includegraphics[width=8cm]{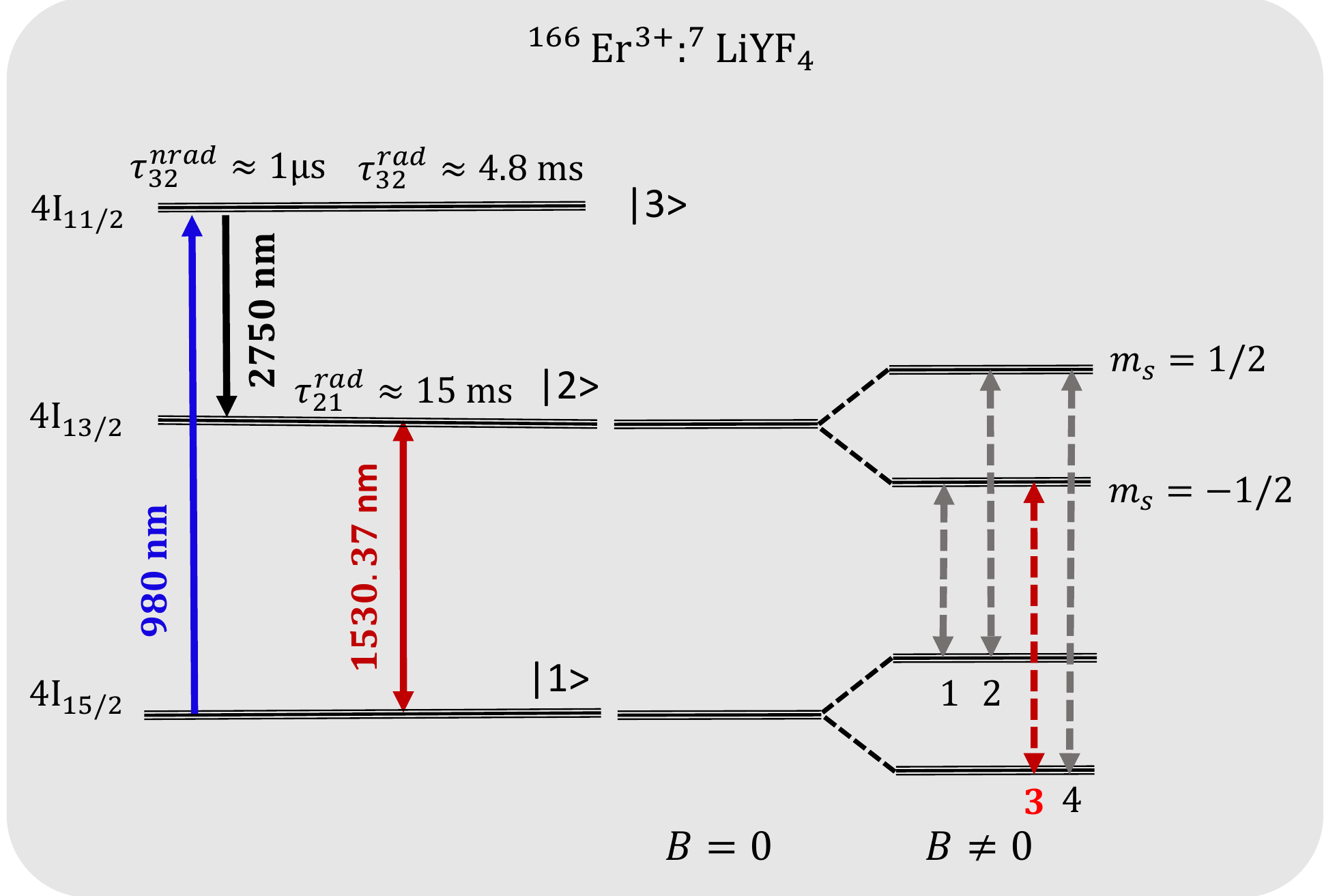}
    \caption{ (Color online) The structure of the energy levels of $^{166}$\erlyf\/ ions. The radiative and non-radiative decay constat is borrowed from Refs. \cite{Ter-Gabrielyan2019,Zyskind1992} }
    \label{Fig1s}
\end{figure}
\par
Rare earth elements are consist of the $15$ lanthanides plus scandium and yttrium. The interesting feature of these elements is the fact that their triply-ionized ions have a partially filled $4f$ shell which is very well shielded by the surrounding $5s^2$ and $5p^6$ electron shells. The resulting inner-shell $4f-4f$ transitions have very narrow line widths spanning a spectrum from the far infrared to the ultraviolet \cite{GuokuiLiu2005}. When doped into a host crystal, the shielding remains effective and the crystal field of the host is merely a weak perturbation of the free ion levels.
\begin{figure}[ht]
    \includegraphics[width=10cm]{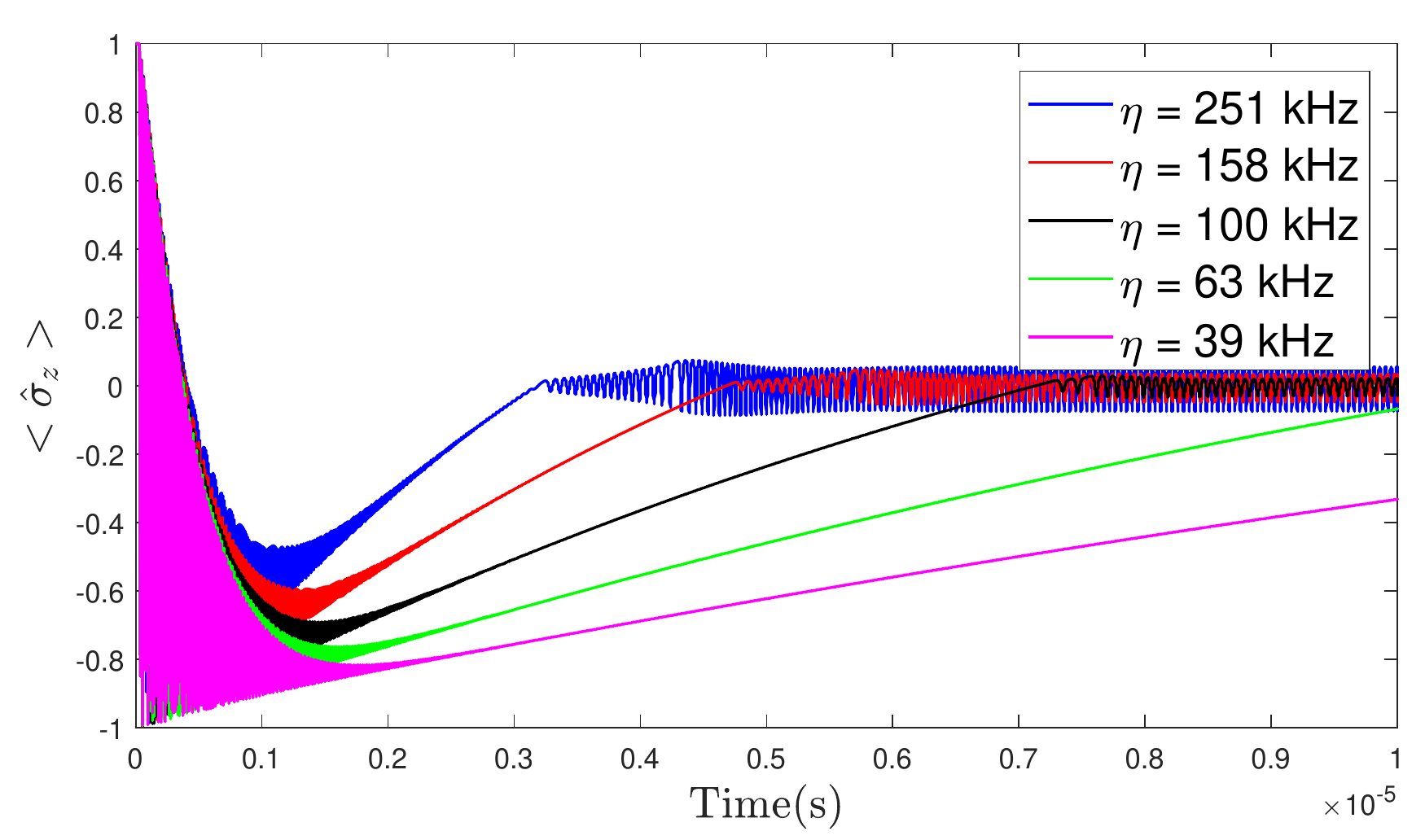}
    \caption{ (Color online) The dynamics of the atomic population $<\hat{\sigma}_z>$ for \pryso. The atomic decay rate is $\gamma=1$ kHz while inhomogeneous broadening is $\chi=100$ kHz. In addition, the cavity linewidth $\kappa=5$ MHz, coupling constant $g=1.4$ kHz, number of atom $N=10^{11}$, and incoherent pumping rate is scanned from $\eta=39$ kHz up to $\eta=251$ kHz in MFT calculation (Eqs. \ref{eq:MFTEq1}-\ref{eq:MFTEq4}). }
    \label{FigS3}
\end{figure}
\par
The energy-level diagram of $^{166}$\erlyf\/ is shown in Fig. \ref{Fig1s}. The electronic configuration of a free Er$^{3+}$ ion is $4f^{11} (n=4,l=3)$, with a $4I$ term. The spin-orbit coupling
splits it into several fine structure levels. An optical transition at the telecom wavelength occurs between the ground state $^{2S+1}L_J = ^4\!I_{15/2}$ and the first excited state $^4\!I_{13/2}$, where S, L, and J are the respective spin, orbital, and total magnetic momenta of the ion. The weak crystal field splits the ground state into eight (J + 1/2) Kramers doublets and the excited state $^4\!I_{13/2}$ being split into seven such double \cite{Bushev2011,Rieger2013}. At cryogenic temperature, only the lowest doublet $Z_1$ is populated, therefore the system can be described as an effective electronic spin with $S = 1/2$. However, erbium has five even isotopes, $^{162}\!Er$, $^{164}\!Er$, $^{166}\!Er$, $^{168}\!Er$, and $^{166}\!Er$, and one odd isotope, $^{167}\!Er$ (natural abundance 22.9\%) with a nuclear spin $I = 7/2$. Therefore, the electronic states of $^{167}\!Er$ with effective spin projection $m_S = \pm1/2$ are additionally split into eight hyperfine levels \cite{Bushev2011}. The even isotopes with nuclear spin $I=0$ has no hyperfine states and therefore are preferred for single mode SRL emission. The narrowest inhomogeneous broadening is measured at $B=197$G to be about $16$ MHz \cite{Kukharchyk2018a,Thiel2011}.

\par
Following Eq. \ref{eq:LLbadCavity}, it is necessary to create at least a moderate population inversion for steady-state SRL emission. In Fig. \ref{Fig1s}, the \dhdhhh transition with $\tau^{rad}_{21}\sim 15$ ms radiative decay time and wavelength of $1530.37$ nm suggested for SRL emission while \pumping transition with $\tau^{nrad}_{32}\sim 1\mu$s is suggested for optical pumping employing $980$ nm light source \cite{Ter-Gabrielyan2019,Zyskind1992}.

\par
Employing Eqs. \ref{eq:MFTEq1}-\ref{eq:MFTEq4} in the main text, one can investigate the dynamics of the atomic population $\langle\hat{\sigma}_z\rangle$ and normalized intracavity photon number $\langle a^{\dagger} a\rangle$ as shown in Fig. \ref{FigS3}. The atomic population $\langle\hat{\sigma}_z\rangle$ dynamics versus different incoherent pumping rate $\eta$ shown in Fig. \ref{FigS3} where provide us the conclusion that having too low pumping rate ($\eta<\gamma$) prevent the SRL emission to reach to the steady-state in a proper time scale.

\end{document}